\documentstyle[sprocl,epsfig]{article}

\newcommand{\AmS}{{\protect\the\textfont2
  A\kern-.1667em\lower.5ex\hbox{M}\kern-.125emS}}
  \newcommand{\bbox}{\lower0.85pt\hbox{$\Box$}}
  \newcommand{\dual}{\mbox{}^{\ast}}
  \newcommand{\kreisl}{\raise0.85pt\hbox{$\scriptstyle\bigcirc$}}
  \newcommand{\dreieck}{\raise0.85pt\hbox{$\scriptstyle\bigtriangledown$}}
  \newcommand{\stern}{\lower0.85pt\hbox{\Large $\star$}}

\def\Journal#1#2#3#4{{#1} {\bf #2}, #3 (#4)}

\def\NPB{{\em Nucl. Phys.} B}
\def\PLB{{\em Phys. Lett.}  B}

\def\PRL{\em Phys. Rev. Lett.}
\def\PRD{{\em Phys. Rev.} D}

\def\be{\begin{equation}}
\def\ee{\end{equation}}
\def\bea{\begin{eqnarray}}
\def\eea{\end{eqnarray}}
\def\bbbone{{\mathchoice {\rm 1\mskip-4mu l} {\rm 1\mskip-4mu l}
{\rm 1\mskip-4.5mu l} {\rm 1\mskip-5mu l}}}

\begin{document}
~
\vspace{-3.0cm}
\begin{flushright}
{\large KANAZAWA-99-01\\
\vskip 3mm
ITEP-TH-5/99}
\end{flushright}
\vspace{0.8cm}

\title{DYNAMICS OF TOPOLOGICAL DEFECTS IN ELECTROWEAK THEORY}

\author{M. N. Chernodub, F. V. Gubarev}

\address{
ITEP, B. Cheremushkinskaya 25, Moscow 117259, Russia\\
E-mails: maxim@heron.itep.ru, Fedor.Gubarev@itep.ru}

\author{E.--M. Ilgenfritz}

\address{Institute for Theoretical Physics, Kanazawa University,\\
Kanazawa 920-1192, Japan \\
E-mail: ilgenfri@hep.s.kanazawa-u.ac.jp}

\author{A. Schiller}

\address{Institut f\"ur Theoretische Physik and NTZ, Universit\"at
Leipzig,\\ D-04109 Leipzig, Germany \\
E-mail: schiller@tph204.physik.uni-leipzig.de }

\maketitle

\abstracts{
\vspace{-1mm}
Embedded defects proposed long ago ($Z$--vortices and Nambu monopoles)
have been successfully searched for in $3D$ equilibrium lattice studies
within the standard model near the electroweak phase transition and the
crossover (which follows it for realistic Higgs mass). Gauge independent
lattice--vortex operators are proposed. Vortex condensation (percolation)
is found to characterize the high--temperature phase. Small vortex clusters
are thermally activated with non--negligible density on the low--temperature
side only at higher Higgs mass, where preliminary evidence supports their
semiclassical nature.
\vspace{-3mm}
}
\section{Introduction}
\vspace{-2mm}
In this contribution some recent work~\cite{CGI,CGIS} has been reviewed
dealing with
the topological signatures in the dimensionally reduced lattice $SU(2)$
Higgs model. This model provides
an effective representation of the
electroweak theory at temperatures near the electroweak scale
in some range of Higgs masses.~\cite{generic,weNPB483}

Lattice studies have shown~\cite{Kajantie,GIS97,Fodor} that the
electroweak theory does not have
the true thermal phase transition (for realistic Higgs mass)
necessary to explain the generation of baryon asymmetry
within the standard model.
Qualitative properties of the model (without extensions)
are still of interest, in order to discuss the high temperature properties of
gauge--matter systems in general and in view of alternative scenarios
of baryogenesis. Topological defects (vortices) and their condensation
(percolation) are expected to play an important role in these contexts.
Therefore, independently of quantitative estimates about the effectiveness
of string--mediated baryogenesis, a first step towards the study of
embedded topological defects seemed to be worthwhile. It has turned
out that, for Higgs masses not much below the endpoint of the phase
transition~\cite{GIS97}
and in the so--called crossover region above, these defects
not only condense in the symmetric phase as they do at lower Higgs mass.
These defects ($Z$--vortices~\cite{Ma83,Na77}) can be thermally generated,
also in the broken phase, as semiclassical objects with non--vanishing density
while they have been shown to be unstable at zero and finite
temperature.~\cite{HolmanVa}
\vspace{-2mm}

\section{The lattice model and defect operators}
\vspace{-1mm}
The $3D$ lattice model is defined by the action
\vspace{-1mm}
\bea
  S &=&  \beta_G \sum_p \big(1 - \frac{1}{2} \mbox{Tr} U_p \big)
   -
  \beta_H \sum_{x,\mu}
  {\mathrm {Re}} \left( \ \phi_x^\dagger U_{x,\mu}\phi_{x+\hat{\mu}}
  \right)
  \nonumber \\
  &&+ \sum_x  \big(  \phi_x^\dagger\phi_x
  + \beta_R \left( \phi_x^\dagger\phi_x -1
  \right)^2 \big)
  \label{eq:S3Dlattice}
\eea
with the 2--component complex isospinor $\phi_x$ for the Higgs field.
The lattice couplings are
related to the continuum parameters of the $3D$ $SU(2)$ Higgs model
$g_3$, $\lambda_3$ and $m_3$. The lattice gauge coupling
$\beta_G=4/(a~g_3^2)$ (with $g_3^2 \approx g_4^2~T$)
gives the lattice spacing in units of temperature, and the hopping parameter
$\beta_H$ substitutes $m_3^2$ driving the transition. The parameter of the
phase transition or the crossover can be translated into a temperature
and a Higgs mass $M_H\approx M_H^*$~\cite{generic,weNPB483} where the parameter
$M_H^*$ is used to parametrize the Higgs self--coupling
\vspace{-2mm}
\be
 \beta_R=\frac{\lambda_3}{g_3^2}
\frac{\beta_H^2}{\beta_G} = \frac{1}{8}
{\left(\frac{M_H^*}{80\ {\mbox {GeV}}}\right)}^2
\frac{\beta_H^2}{\beta_G}\, .
\label{MH*}
\ee

The $Z$--vortex~\cite{Ma83,Na77} corresponds to the
  Abrikosov-Nielsen-Olesen~\cite{ANO} vortex solution related to the
  Abelian subgroup of $SU(2)$ and embedded into the $SU(2)$ gauge field
  of the electroweak theory.
We use gauge invariant lattice definitions for
$Z$--vortices and Nambu mono\-poles as extended topological objects
of size $k a$.
The construction  for elementary ($k$=1) defects proceeds as follows.
A composite adjoint unit vector field
$n_x = n^a_x \sigma^a$,
$n^a_x = - ({\phi^+_x \sigma^a \phi_x})/({\phi^+_x \phi_x})$ is introduced
which allows to define the gauge invariant
flux ${\bar \theta}_p$ through the pla\-quet\-te $p=\{x,\mu\nu\}$,
\vspace{-2mm}
\be
{\bar \theta}_p =  \arg \Bigl( {\mathrm {Tr}}
\left[(\bbbone + n_x) V_{x,\mu} V_{x +\hat\mu,\nu}
V^+_{x + \hat\nu,\mu} V^+_{x,\nu} \right]\Bigr)
\label{AP}
\ee
via the projected links
$V_{x,\mu}(U,n) =  U_{x,\mu} + n_x U_{x,\mu} n_{x + \hat\mu}$.
The plaquette angle $\chi_p$ is constructed with the help of the
Abelian link angles
$\chi_{x,\mu}=\arg\left(\phi^+_x V_{x,\mu} \phi_{x + \hat\mu}\right)$
as usual:
$
\chi_{p} = \chi_{x,\mu} + \chi_{x +\hat\mu,\nu} -
\chi_{x + \hat\nu,\mu} - \chi_{x,\nu}$.
The $Z$--vorticity number $\sigma_p$ of plaquette $p$ and
the monopole charge $j_c$ carried by the cube $c$
(defined in terms of the fluxes
(\ref{AP}) penetrating the surface $\partial c$)
are given by
\be
\sigma_p = \frac{1}{2\pi} \Bigl( \chi_p - {\bar \theta}_p \Bigr) \,, \ \ \
j_c = - \frac{1}{2\pi} \sum_{p \in \partial c}
{\bar \theta}_p\,.
\label{SigmaNjN}
\ee
A $Z$--vortex is formed by links $l=\{x,\rho\}$ of the dual
lattice ($l$ dual to $p$) which  carry
a non--zero vortex number: $\dual \sigma_{x,\rho} =
\varepsilon_{\rho\mu\nu} \sigma_{x,\mu\nu} \slash 2$.
$Z$--vortex trajectories
are either closed or begin/end on Nambu (anti-) monopoles:
$
\sum^3_{\mu=1} (\dual \sigma_{x-\hat\mu,\mu} - \dual \sigma_{x,\mu})
= \dual j_x$.
Extended monopoles (vortices) on $k^3$ cubes ($k^2$ pla\-quet\-tes)
are constructed
analogously
replacing the elementary
plaq\-uet\-tes in terms of $V_{x,\mu}$
by Wilson loops of corresponding size.

We call a vortex cluster a set of connected dual links carrying non--zero
vorticity (vortex trajectories). A bond percolation algorithm (known from
cluster algorithms for spin models) has been used to separate the various
disconnected $Z$--vortex clusters that coexist in a lattice configuration.

We have measured the total densities
$\rho_m = \sum_c |j_c|/L^3$ of Nambu monopoles and
$\rho_v = \sum_p |\sigma_p|/(3~L^3)$ of vortex links
as well as the
percolation probability of $Z$--vortex trajectories
$C = \lim_{r \to \infty} C(r)$ derived from the cluster
correlation function
$
C(r) = {\sum_{x,y,i}
\delta_{x \in \dual \sigma^{(i)}} \,\delta_{y \in \dual \sigma^{(i)}}
\cdot \delta (|x-y|-r)}
/{\sum_{x,y} \delta (|x-y|-r)}$.
A cluster $\dual \sigma^{(i)}$ contributes to the correlator if the vortex
lines pass through both points $x$ and $y$.
The average number of $Z$--vortex clusters and the average number
of dual $Z$--vortex links per cluster have been measured to characterize
the structural change near the percolation transition across
the electroweak crossover.
\vspace{-3mm}

\section{Monte Carlo results}
\vspace{-1mm}
\subsection{At thermal first order phase transition}
\begin{figure}[!htb]
 \vspace{-2mm}
 \begin{minipage}{7.5cm}
 \begin{center}
 \epsfig{file=
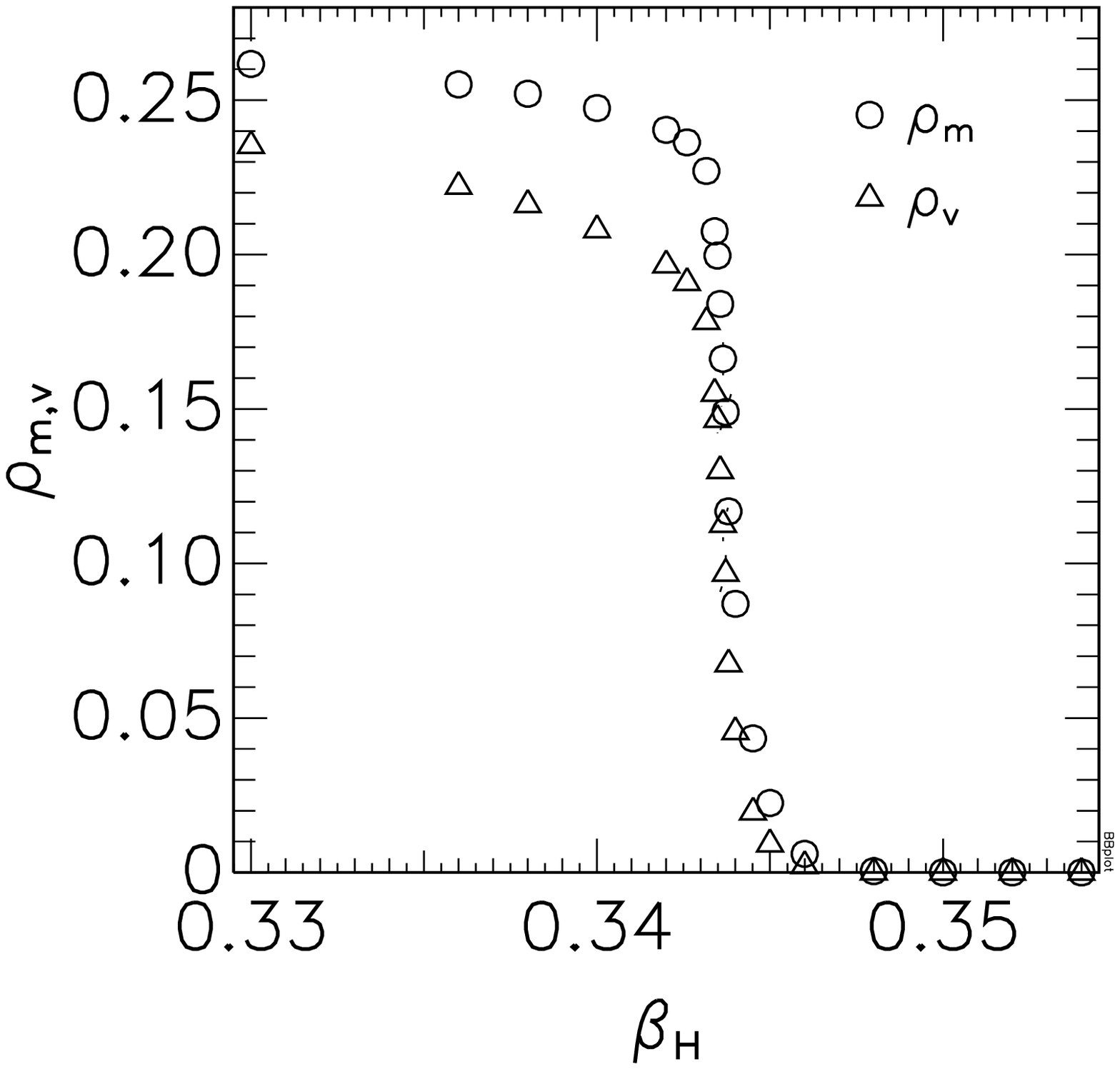,width=3.6cm,height=3.6cm}
 \epsfig{file=
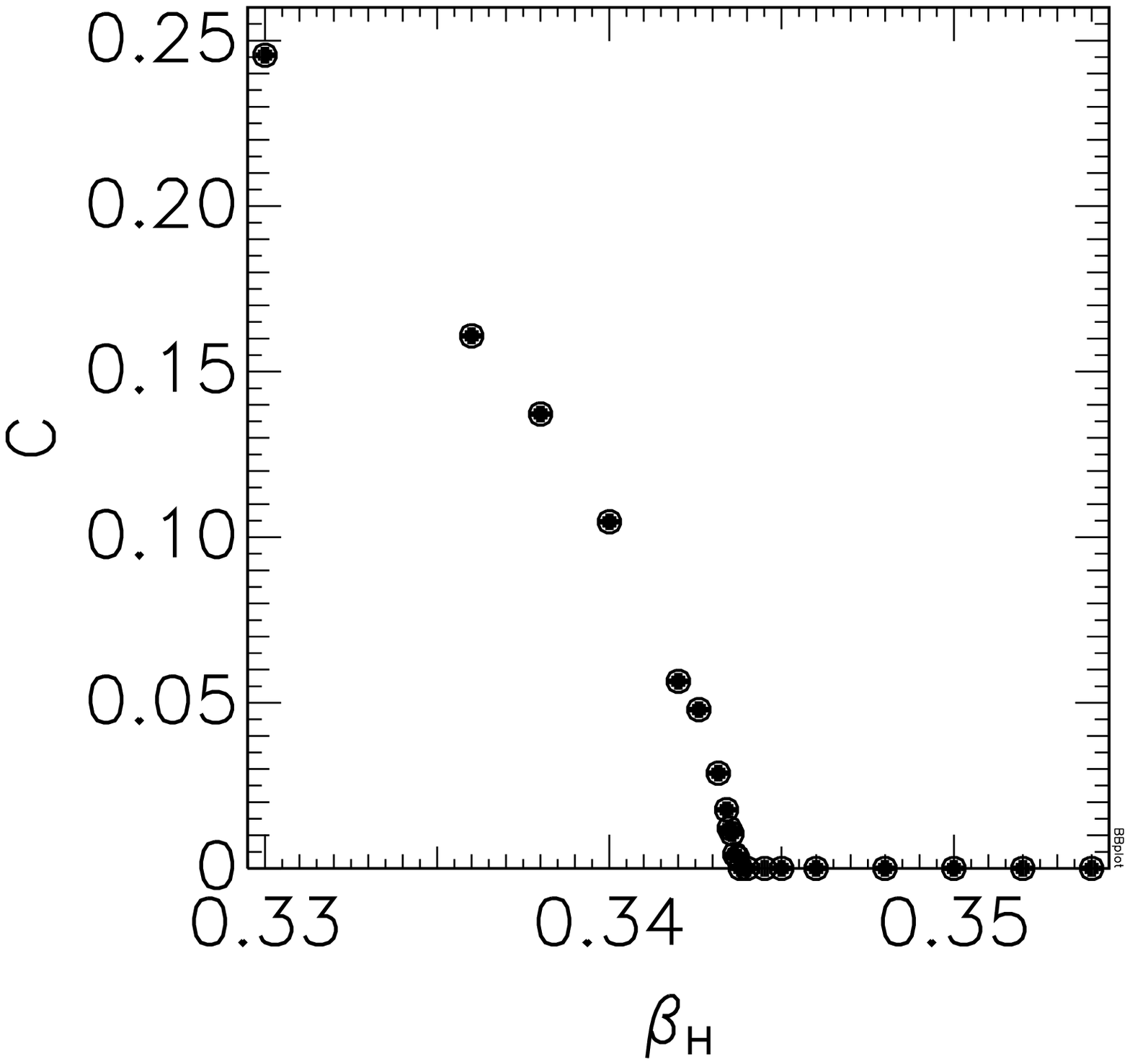,width=3.6cm,height=3.6cm}
 \end{center}
 \end{minipage}
 \begin{minipage}{4.3cm}
 \begin{center}
 \caption{
 Densities of elementary Nambu monopoles $\rho_m$ and
 $Z$--vortices $\rho_v$ (left)
 and percolation probability $C$ (right) vs. $\beta_H$
 at $M_H^*= 70$~GeV and $\beta_G = 12$.}
 \label{fig:1}
 \end{center}
 \end{minipage}
 \vspace{-4mm}
\end{figure}

\vspace{-1mm}
We have scanned the phase transition with elementary
defects at $M_H^*=30$ GeV (strong first order) and 70 GeV
(weak first order) for $\beta_G=12$ and lattice volume $16^3$.
At the lower Higgs mass we have observed a discontinuity of the densities
$\rho_{m,v}$ jumping to zero at the critical temperature $T_c$.
The percolation probability $C$ has a
finite jump to zero at $T_c$.
The same study  near the
endpoint~\cite{GIS97} of the first order
transition is summarized in Fig.~\ref{fig:1}
for increasing $\beta_H$ (decreasing temperature).
Now the percolation probability continuously vanishes towards
$\beta_{Hc}=0.34355$.
There are inflection points of the densities $\rho_m$ and $\rho_v$ at this
value of $\beta_H$ where the corresponding objects are approximately half as
abundant compared to the symmetric phase. For $\beta_H > \beta_{Hc}$
the densities decrease exponentially.
Within our accuracy and using the mentioned percolation definition
the critical temperature $T_c$
and the percolation temperature $T_{\mathrm{perc}}$
coincide in this model.
\vspace{-2mm}

\subsection{In the crossover region}
\vspace{-1mm}

\begin{figure}[!htb]
\vspace{-3mm}
\begin{minipage}{7.2cm}
  \begin{center}
  \epsfig{file=
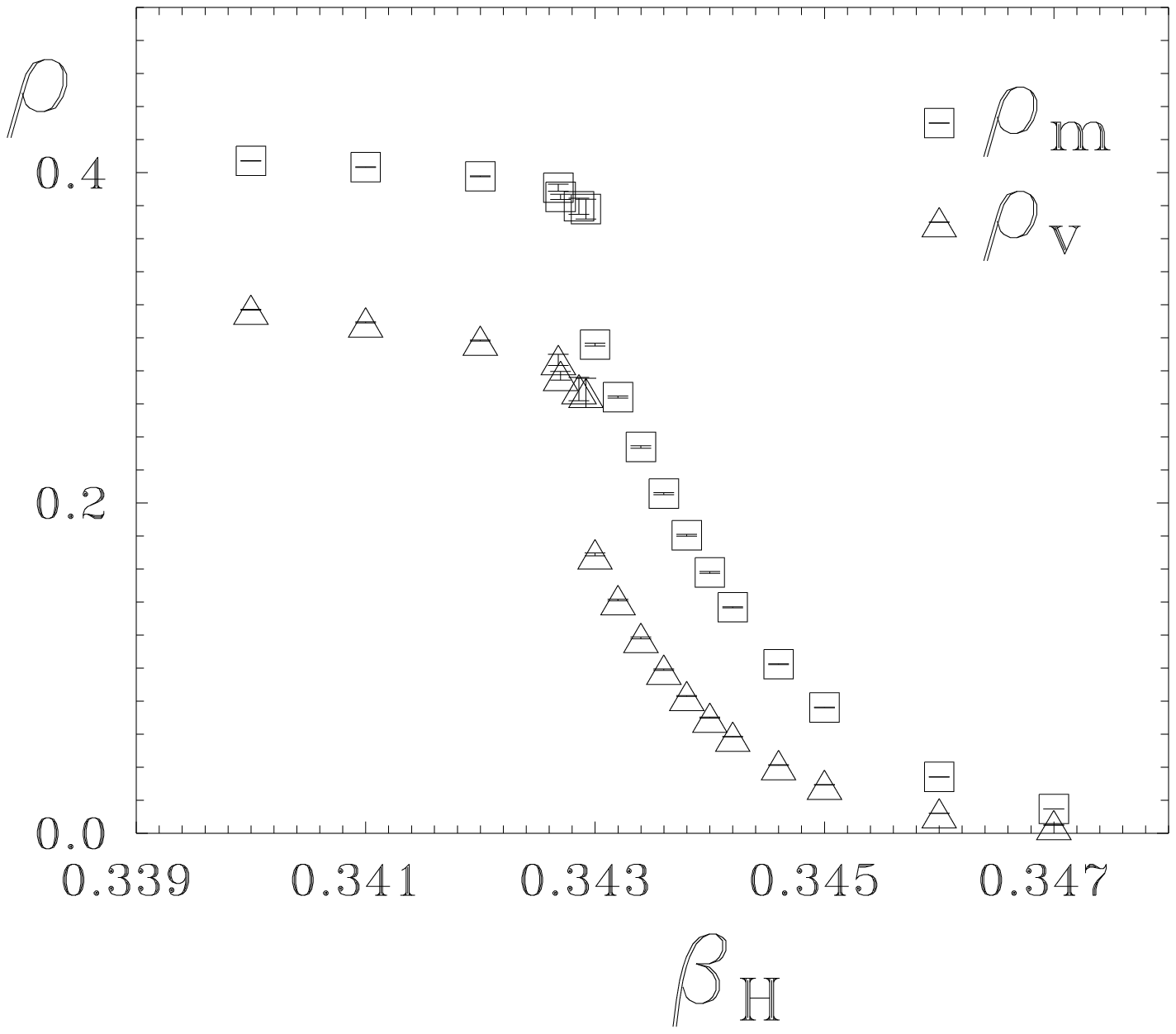,width=3.4cm,height=3.6cm}
  \epsfig{file=
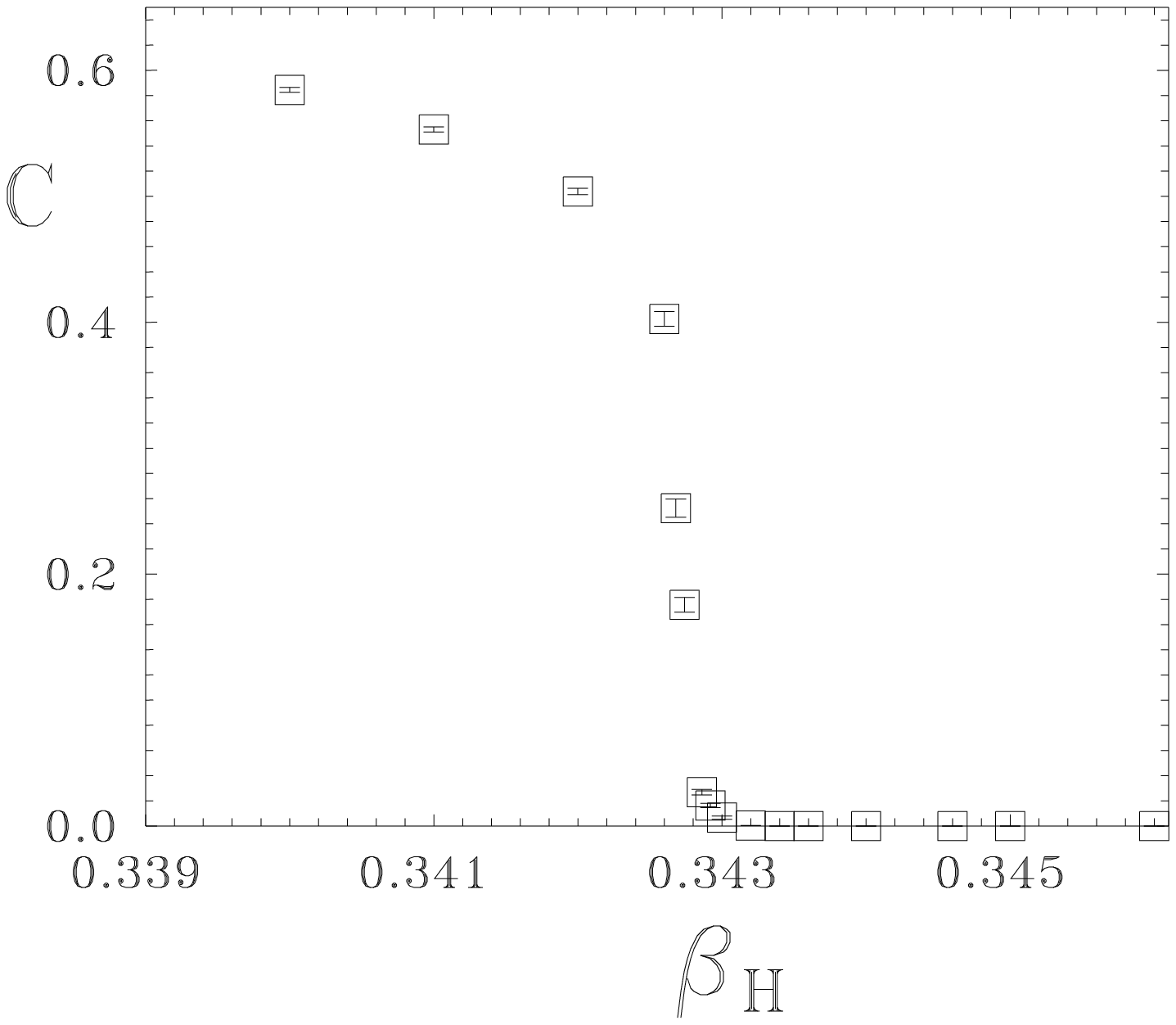,width=3.4cm,height=3.6cm}
  \vspace{-3mm}
  \caption{
  Same as Fig.~\protect\ref{fig:1} for defects of size $k$=2
  at $M_H^*=100$~GeV, $\beta_G = 16$ and lattice $32^3$.}
  \label{fig:2}
  \end{center}
\end{minipage}
\begin{minipage}{4mm}
\end{minipage}
\begin{minipage}{4.4cm}
  \begin{center}
  \vspace{ 1.0mm}
  \epsfig{file=
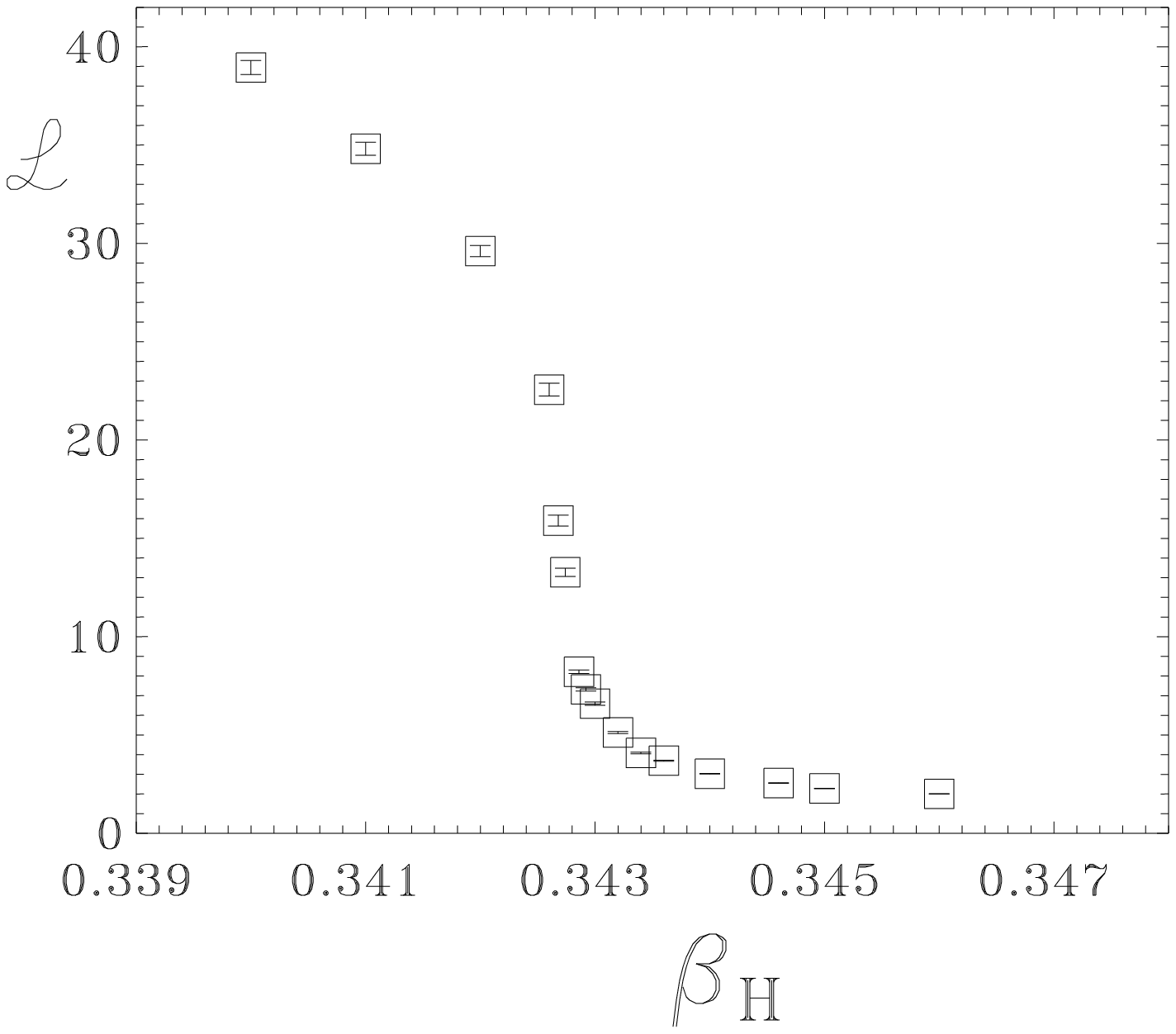,width=3.4cm,height=3.5cm}
  \vspace{-4mm}
  \caption{Average length ${\cal L}$ per cluster (vortex size $k=2$)
  for the same ensemble as in Fig.~\protect\ref{fig:2}.}
  \label{fig:3}
  \end{center}
\end{minipage}
\vspace{-4mm}
\end{figure}

{}From a phenomenological point of view the
crossover region, investigated in
our studies at $M^*_H=100$ GeV, is more interesting because it is not
excluded by experimental evidence.
The Monte Carlo results presented in Figs.~\ref{fig:2},\ref{fig:3}
show the existence of a network of $Z$--vortices on the
high--temperature side of the crossover, with finite probability
of percolating (at $T > T_{\mathrm{perc}}$),
while only smaller clusters occur below $T_{\mathrm{perc}}$,
however much more abundantly than at lower Higgs mass. Some of these
clusters are closed, others open (with Nambu monopoles at the ends).
We have checked that there is an universal percolation temperature
for vortices of extension $k~a$ at respective $\beta_G^{(k)}=
k~\beta_G^{(1)}$ keeping the physical volume fixed.
The estimate is $T^{\mathrm{perc}}=170$ or
130 GeV (without or with $t$--quarks) with a Higgs mass
$M_H=94$ or 103 GeV.
We note that the crossover/percolation transition is also accompanied
by interesting physics in so far as the spectral evolution in
various  channels is characterized by strong mixing of gauge and scalar
degrees of freedom.~\cite{EPJC99}

The physical implications of the decay of the percolating network
of $Z$--strings into many small clusters seem to be interesting and
worth to be studied more in detail. For a cosmological context the
kinetics of this phase transition is interesting. It will differ
from the bubble dominated transition of the strong first order
transition and needs real time simulations.

\begin{figure}[!htb]
\vspace{-4mm}
\begin{minipage}{5.5cm}
\begin{center}
\vspace{-7mm}
\epsfig{file=
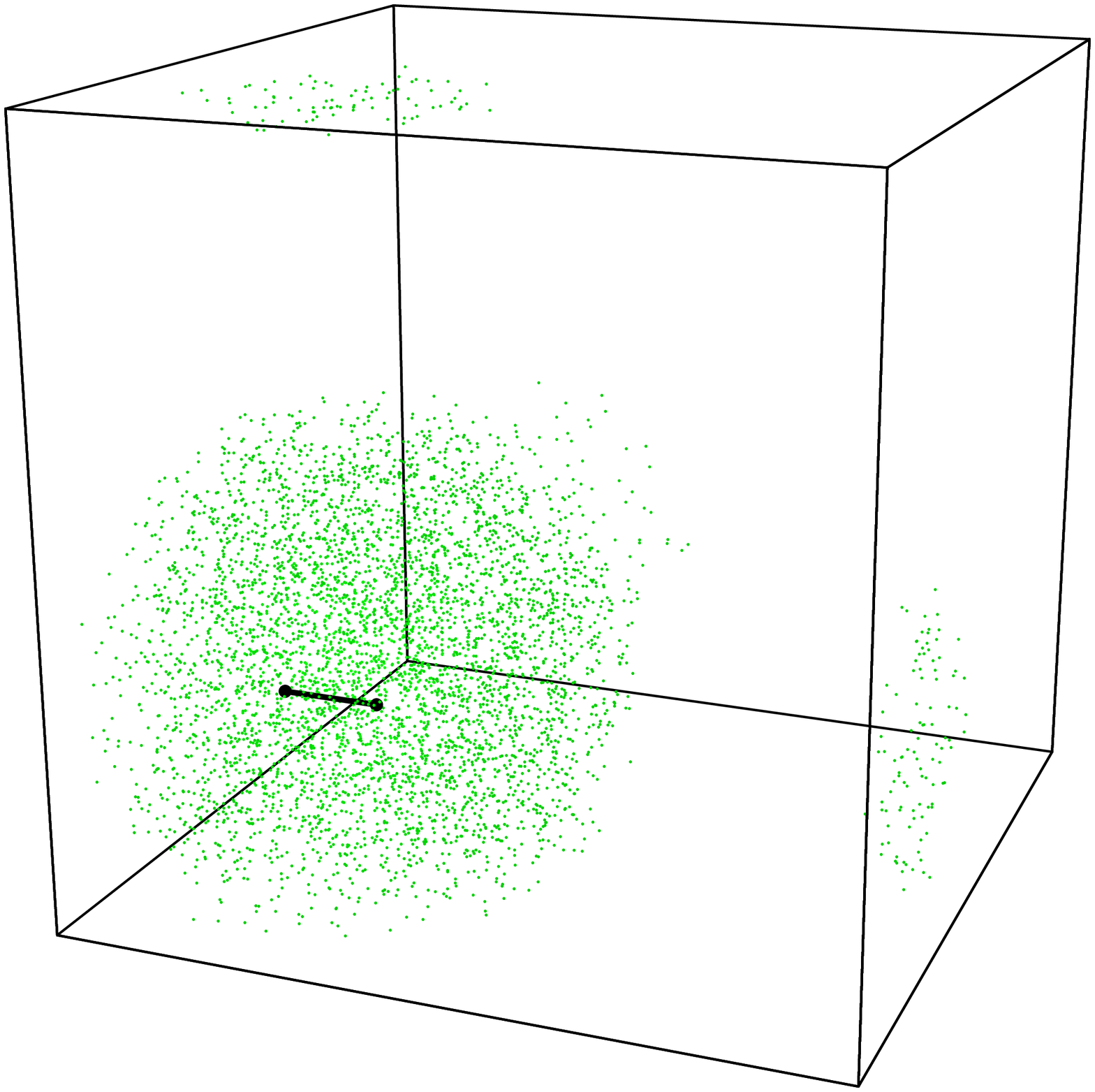,width=5.00cm}
\vspace{-11mm}
\caption{
Classical sphaleron as a Nambu monopolium bound by a $Z$-vortex
string. The clouds show the suppression in Higgs field modulus.}
\label{fig:4}
\end{center}
\end{minipage}
\begin{minipage}{7mm}
\end{minipage}
\begin{minipage}{5.8cm}
\vspace{2mm}
\begin{center}
\epsfig{file=
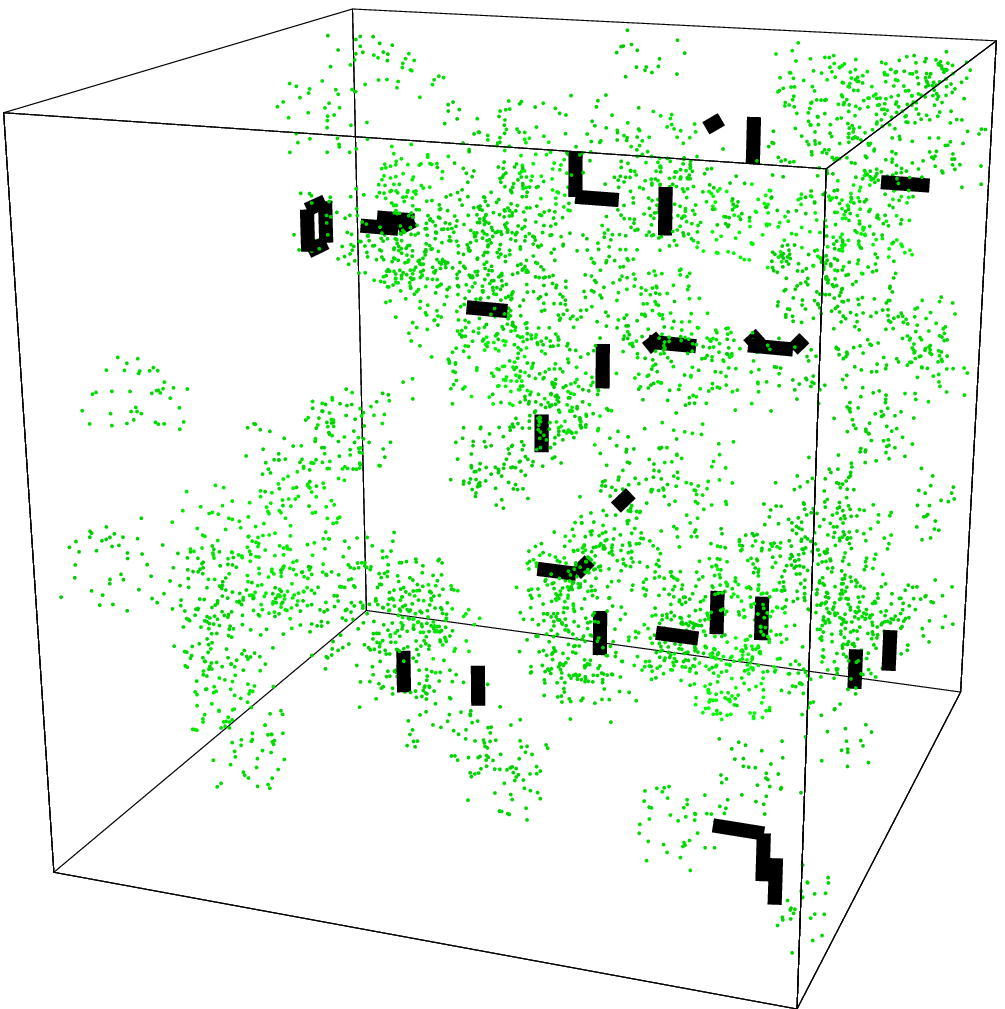,width=4.7cm}
\vspace{-2mm}
\caption{
Clusters of extended ($k=3$) $Z$-vortices just below
$T_{\mathrm{perc}}$.
Nambu monopoles are not shown. The Higgs modulus is visualized as in
 Fig.~\protect\ref{fig:4}.}
\label{fig:6}
\end{center}
\end{minipage}
\vspace{-4mm}
\end{figure}

We propose to identify some fraction of the density $\rho_v$ on the
``broken phase'' side of the crossover as the density of sphalerons.
This conjecture is supported by the signature
of a
{\it classical} lattice sphaleron with respect to the new ($Z$--vortex and
Nambu monopole) degrees of freedom.
Fig.~\ref{fig:4} shows one of the solutions~\cite{vanBaal} with a Nambu
monopole--antimonopole pair in the center.
 Note also that the average cluster size is ${\cal L}=2$
(Fig.~\ref{fig:3}).

\begin{figure}[!htb]
\vspace{-3mm}
\begin{minipage}{7.5cm}
\begin{center}
\epsfig{file=
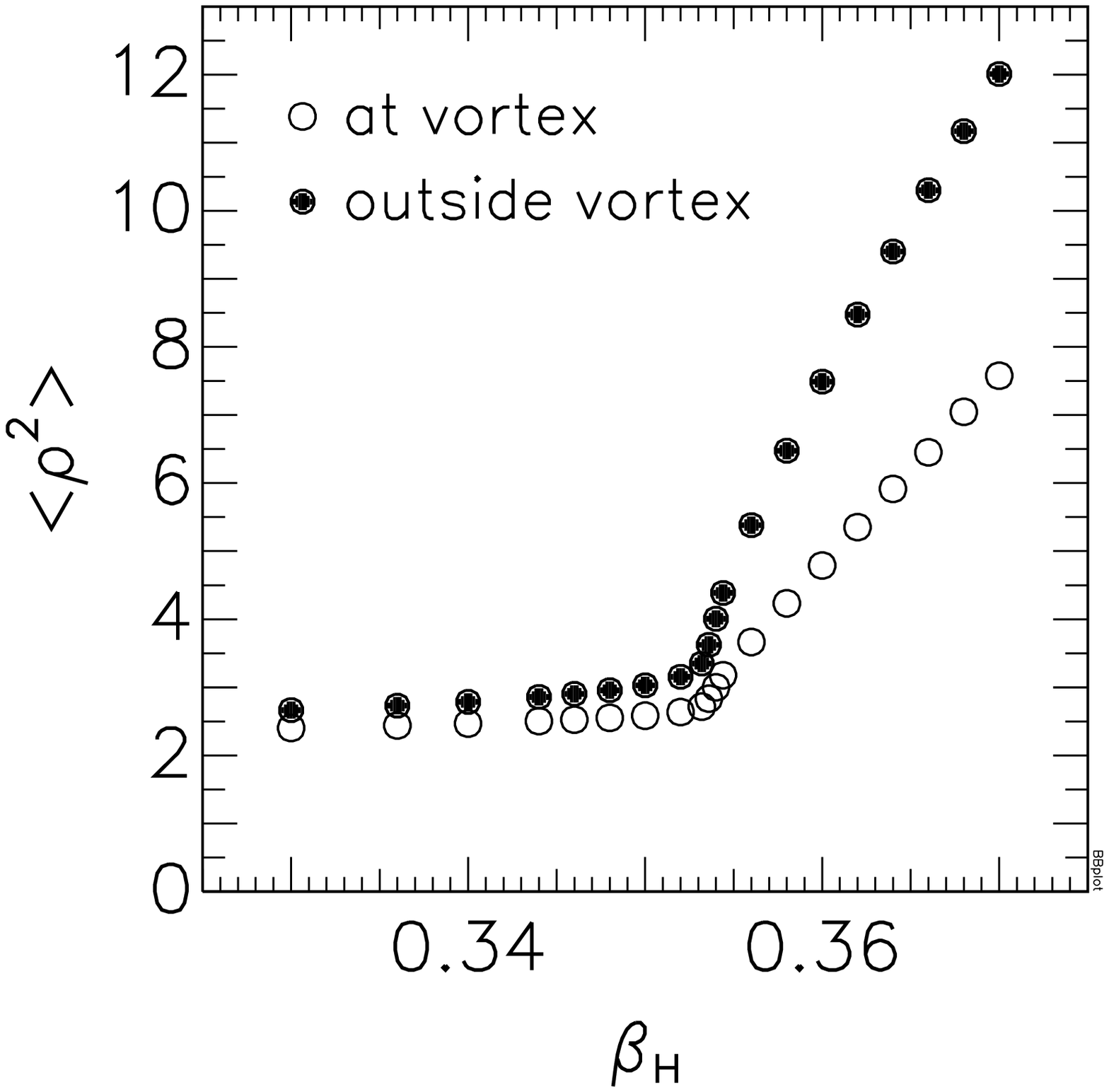,width=3.65cm,height=3.7cm}
\epsfig{file=
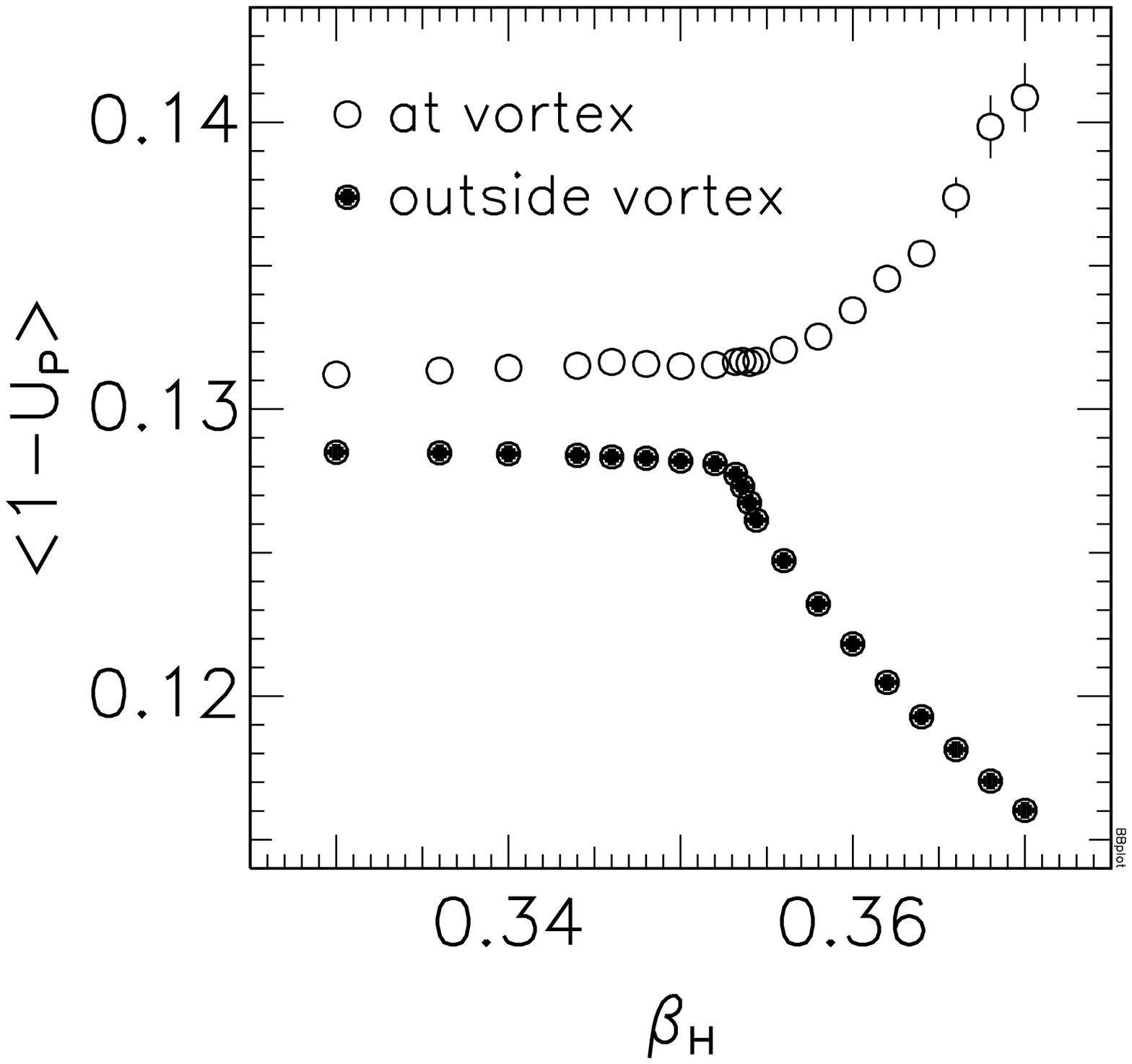,width=3.65cm,height=3.7cm}
\end{center}
\end{minipage}
\begin{minipage}{4.3cm}
\begin{center}
\caption{
Average squared Higgs field modulus (left) and gauge field energy (right)
inside and outside an elementary ($k$=1) $Z$--vortex on both sides of the
percolation transition, ($M_H=100$~GeV, $\beta_G=8$).}
\label{fig:5}
\end{center}
\end{minipage}
\vspace{-5mm}
\end{figure}

Why do we believe that the lattice defect operators (\ref{SigmaNjN})
detect not just lattice artifacts?
Encouraging  although preliminary quantitative
information is provided by investigations of the continuum limit
(this requires the calculation of the density of extended defects
having various sizes on finer lattices)
and by measuring local averages of gauge field action and
Higgs field near the $Z$--vortex soul.
Fig.~\ref{fig:5}
shows this for elementary
vortices (non--zero {\it vs.} zero vorticity plaquettes) for a rather coarse
lattice. When vortices are condensed in the symmetric phase, their local
averages differ only slightly from no--vortex averages.
Local action and Higgs field are distinctly different from the bulk
in the broken phase.
The correlations between the position of the vortices and decreasing
Higgs field values (larger cloud densities)
are exemplified in Fig.~\ref{fig:6}, where we show a snapshot using $k=3$
vortices (blocked from a $48^3$ lattice)
for $\beta_G=24$ at a temperature slightly below $T_{\mathrm{perc}}$
($\beta_H=0.3630$).
This Figure supports the semiclassical nature of the defects.
\vspace{-2mm}

\section{Outlook}
\vspace{-1mm}
For the next future, the systematic exploration of the continuum limit
with lattices of comparable physical size has to be completed.
For the broken phase we expect to obtain well--defined
size distribution and internal profile of the embedded $Z$--vortices while,
for the condensed phase, this will probably not be possible.

We plan to extend our considerations to more realistic models with
$\theta_W\ne0$. In order to clarify the connection between the
dynamics of vortices with the evolution of Chern--Simons number, the
$3$--dimensional studies presented here have to be complemented by
Euclidean and real--time simulations. This seems to be an interesting
piece of physics whether or not it finally leads to a viable
mechanism of baryogenesis.  \vspace{-2mm}

\section*{Acknowledgments}
\vspace{-2mm}
M.~N.~Ch. and F.~V.~G. were partially supported by the grants
INTAS-96-370, INTAS-RFBR-95-0681, RFBR-96-02-17230a and
RFBR-96-15-96740. M.~N.~Ch. was also supported by
the INTAS Grant 96-0457 (ICFPM program).

\end{document}